\documentstyle[epsf,twocolumn,seceq,short]{jpsj}

\def\runtitle{Filling control of the pyrochlore oxide Y$_{2}$Ir$_{2}$O$_{7}$}
\def\runauthor{Hideto {\sc Fukazawa} and Yoshiteru {\sc Maeno}}

\title
{
Filling control of the pyrochlore oxide Y$_{2}$Ir$_{2}$O$_{7}$
}

\author
{ 
Hideto {\sc Fukazawa}$^{1,}$\footnote{E-mail address: hideto@scphys.kyoto-u.ac.jp}
and Yoshiteru {\sc Maeno}$^{1,2}$
}

\inst
{
$^1$Department of Physics, Kyoto University, Kyoto 606-8502\\
$^2$International Innovation Center, Kyoto University, Kyoto 606-8501\\
}

\recdate
{
\today
}

\kword
{
Y$_{2}$Ir$_{2}$O$_{7}$, pyrochlore, Mott insulator, metal-insulator transition, 
iridate, filling control
}

\begin{document}
\sloppy
\maketitle

Recently pyrochlore iridates, $R_{2}$Ir$_{2}$O$_{7}$ ($R$ = Y and lanthanides), 
have provided additional research interests 
in geometrically frustrated magnetic systems. \cite{Yan1,Tai1,Bra1} 
These materials were reported as early as about 30 years ago 
but were little studied. \cite{Sub1} 
Kennedy {\it et al.} \cite{Ken1} studied the crystal structure of the materials, 
but they have not reported the low-temperature properties, 
which are essential for the characterization of the frustrated magnetic systems. 

The magnetic frustration arises because of the pyrochlore structure in which $R$ and 
Ir sub-lattices form individual networks of linked tetrahedra. 
Systematic variations of the physical properties have been 
investigated by changing the elements of the $R$ site. \cite{Yan1} 
Furthermore, one may expect that each Ir ion has a quantum spin $S$ = 1/2 
if the 5 $5d$ electrons in the $t_{2g}$ orbitals are localized 
in the presence of strong electronic correlation. 
Although a quantum spin liquid state is theoretically expected 
for antiferromagnetic Heisenberg pyrochlore magnet, \cite{Can1} 
there have been few candidates of the actual materials studied up to now. 
Besides, Fujimoto \cite{Fuj1} theoretically showed for a hole-doped pyrochlore Mott-insulator 
based on an $s$-electron system that the electronic specific-heat coefficient $\gamma$ 
exhibits a divergent behavior near the boundary of metal-insulator transition. 
These urge us to study the $S = 1/2$ pyrochlore system, pyrochlore iridates. 

Of the pyrochlore iridates, Y$_{2}$Ir$_{2}$O$_{7}$ serves as a reference material, 
since it does not possess a magnetic rare-earth element on the $R$ site and 
its physical properties are comparable with those of its isomorphs Y$_{2}$Mo$_{2}$O$_{7}$ 
\cite{Raj1} and Y$_{2}$Ru$_{2}$O$_{7}$ \cite{Yos1}, based on $S = 1$. 

Y$_{2}$Ir$_{2}$O$_{7}$  exhibits non-metallic behavior down to 4.2 K and exhibits 
quite a small ferromagnetic (FM) component below $T_{{\rm m}}$ = 170 K, though it has not been 
well understood whether such FM component is intrinsic or not. \cite{Yan1,Tai1} 
In spite of the non-metallic behavior, it has been reported by Taira {\it et al.} \cite{Tai1} 
that the  $\gamma$ is finite, 4.1(3) mJ/K$^{2}$mol-Ir, 
suggesting the existence of the Fermi surface. 
We will discuss the origins of the finite  $\gamma$ 
and the FM component of Y$_{2}$Ir$_{2}$O$_{7}$. 
It is vital to investigate the filling control of Y$_{2}$Ir$_{2}$O$_{7}$ 
in order to search for the metallic phase adjacent to the strongly correlated nonmetallic phase. 
We succeeded in synthesizing hole-doped material, 
namely Y$_{2-x}$Ca$_{x}$Ir$_{2}$O$_{7}$ ($x$ = 0.2, 0.3, 0.4 and 0.6). 
As the main purpose of this note, we will show the metal/non-metal (M/NM) crossover and 
the enhancement of the $\gamma$ in Y$_{2-x}$Ca$_{x}$Ir$_{2}$O$_{7}$. 

We used polycrystals synthesized by the conventional solid-state-reaction method. \cite{Yan1} 
We measured the resistivity by a standard four-probe method below 300 K and 
the specific heat by a relaxation method between 1.8 and 300 K (Quantum Design, PPMS). 
We investigated the dc magnetization with a SQUID magnetometer 
(Quantum Design, MPMS$_{\rm 5S}$) between 1.8 and 350 K. 

\begin{figure}
\leavevmode
\epsfxsize=70mm
\epsfbox{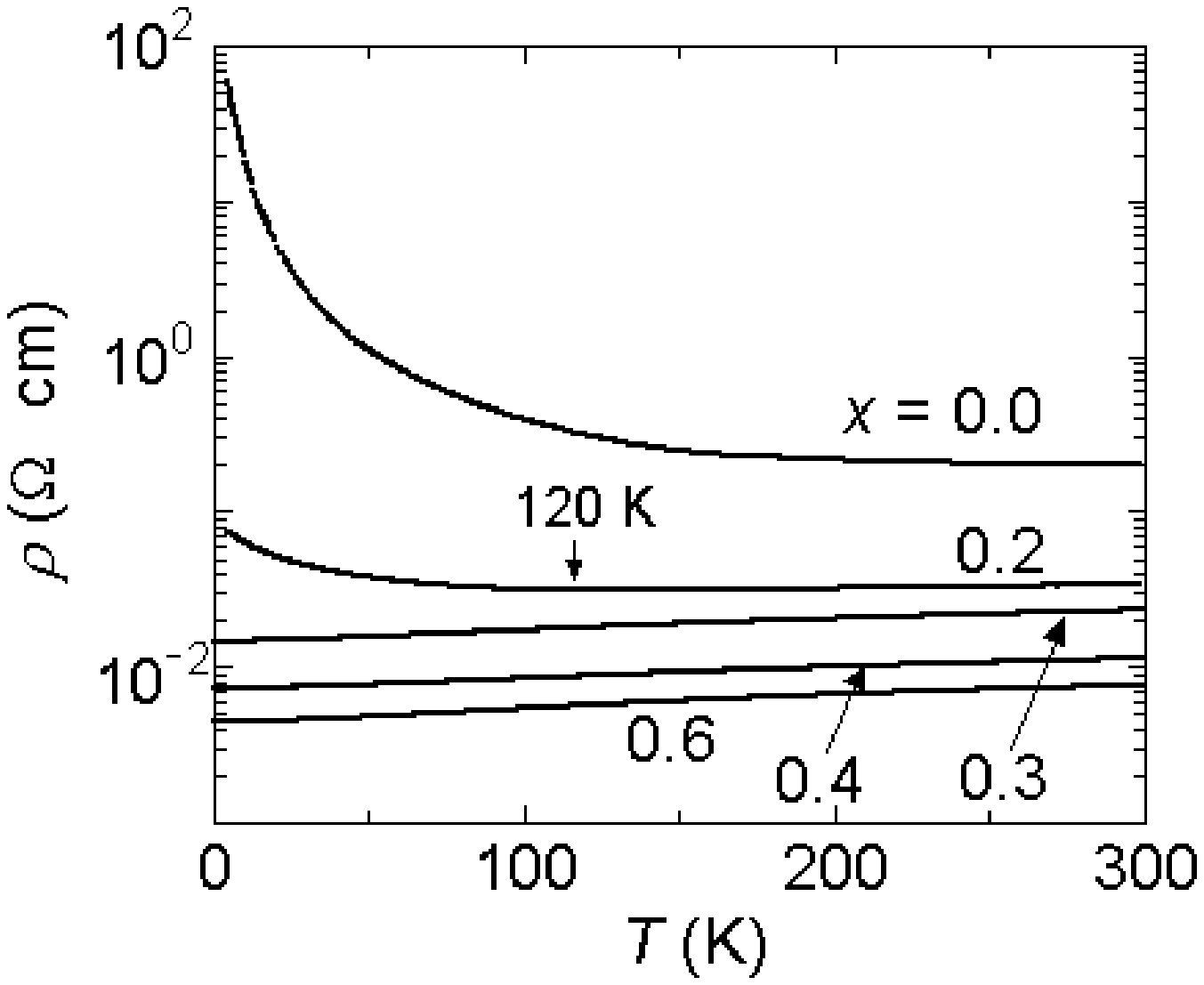}
\caption{Resistivity of Y$_{2-x}$Ca$_{x}$Ir$_{2}$O$_{7}$. }
\label{fig1}
\end{figure}

The lattice parameters at room temperature depend little on $x$ 
($a = 10.184(2) {\rm \AA}$, cubic) except for $x$ = 0.0 ($a = 10.176(1) {\rm \AA}$, cubic). 
We note that transport properties at room temperature also change 
from non-metallic to metallic between $x$ = 0.0 and 0.2. 

In Fig. 1, we plot the resistivities of Y$_{2-x}$Ca$_{x}$Ir$_{2}$O$_{7}$. 
Y$_{2}$Ir$_{2}$O$_{7}$ exhibits non-metallic behavior, as we previously reported. \cite{Yan1} 
For $x$ = 0.2, it exhibits metallic behavior down to about 120 K, 
and non-metallic behavior at lower temperature. 
For $x$ = 0.3, 0.4 and 0.6, it exhibits metallic behavior 
at least down to 0.3 K; however, no sign of superconductivity has been observed. 

In Fig. 2, we plot the dc magnetic susceptibilities 
$M(T)/H \equiv \chi (T)$ ($\it{\mu}_{\rm{0}}\it{H}$ = 1 T). 
Very small FM component, amounting to $4 \times 10^{-3}$ of the saturated moment of $S = 1/2$ 
spins, is observed for $x$ = 0.0 ($T_{{\rm m}}$ = 170 K) and 0.2 ($T_{{\rm m}}$ = 100 K). 
The $T_{{\rm m}}$ of Y$_{2}$Ir$_{2}$O$_{7}$ is consistent with the value 
previously reported by us \cite{Yan1} and by Taira {\it et al}. \cite{Tai1} 
The FM component was not observed above $x$ = 0.3. 
Thus, the magnetic ground state appears to be correlated with non-metallic ground state. 
We obtained the effective spin $S_{{\rm eff}}$ = 0.07(1) from the Curie-Weiss fitting 
($\chi (T) = \chi_{0} + 
\frac{4\mu_{{\rm B}}^{2} S_{{\rm eff}}(S_{{\rm eff}} +1)}{3k_{{\rm B}}(T- \theta_{{\rm CW}})}$) 
for Y$_{2}$Ir$_{2}$O$_{7}$ above $T_{{\rm m}}$. 
It corresponds to only 14(2)\% of the expected spin $S$ = 1/2. 

Small magnetic moment of Y$_{2}$Ir$_{2}$O$_{7}$ below $T_{\rm m}$ 
may be due to either spin-glass ordering or canted antiferromagnetic ordering. 
Since no anomaly was observed at around $T_{\rm m}$ in specific heat (data not shown), 
the FM component is attributable to spin-glass ordering 
as previously reported in other pyrochlores, such as Y$_{2}$Mo$_{2}$O$_{7}$ \cite{Raj1} 
and Y$_{2}$Ru$_{2}$O$_{7}$ \cite{Yos1}$\!\!\!$ . 

We should note that an additional steep increase of magnetization was observed 
for all the materials with decreasing temperature below about 15 K. 
Since no change in $\rho {\rm (}T{\rm )}$ was observed below this temperature, 
the increase is attributable to magnetic impurities or Ir spins at grain boundaries. 

\begin{figure}
\leavevmode
\epsfxsize=70mm
\epsfbox{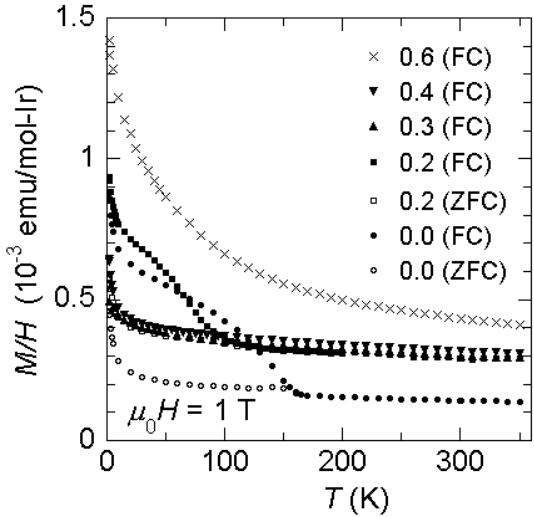}
\caption{Magnetic susceptibilities of Y$_{2-x}$Ca$_{x}$Ir$_{2}$O$_{7}$. 
Solid symbols denote the data obtained after field-cooling. 
Open symbols denote the data obtained after zero-field-cooling. }
\label{fig2}
\end{figure}

In Fig. 3, we plot the specific heat divided by temperature, $C_{P}(T)/T$, against $T^{2}$. 
Solid line for $x$ = 0.0 denotes the quadratic fitting, 
$C_{P}/T = \gamma + \beta T^{2}$, between 14 and 20 K. 
Below 14 K, it is difficult to perform a valid quadratic fitting 
because of additional increase of $C_{P}/T$ with decreasing $T$. \cite{Tai2} 
This increase cannot be explained in terms of the localization effect, 
since  in a such case $C_{P}/T$ should exhibit linear behavior (the Debye $T^{3}$ term) 
with finite intercept $\gamma$ at least up to 20 K. 
We speculate that the origin of the increase is attributable to a magnetic contribution, 
since it appears to correspond to the steep increase of $\chi(T)$ below 15 K. 

Although $C_{P}/T$ of Y$_{2}$Ir$_{2}$O$_{7}$ at 1.8 K is 5.8(2) mJ/K$^{2}$mol-Ir, 
we may consider that the $\gamma$, which is equal to 0.0(5) mJ/K$^{2}$mol-Ir, 
is the intrinsic $\gamma$. 
This value strongly suggests that Y$_{2}$Ir$_{2}$O$_{7}$ is a Mott insulator. 

In the inset of Fig. 3, we show the $\gamma$ against the substitution content $x$. 
The $\gamma$ is obtained from the quadratic fitting between 14 and 20 K. 
We note that the Debye temperature is ${\it \Theta}_{{\rm D}}$ = 400(10) K 
and depends little on $x$. 
Once additional holes are introduced into the half-filled $t_{2g}$ band, 
the finite density of states at the Fermi level appears. 
The $\gamma$ for $x$ = 0.2 is 8.1(5) mJ/K$^{2}$mol-Ir much greater than the value, 1(1) 
mJ/K$^{2}$mol-Ru, for the corresponding material Y$_{1.8}$Bi$_{0.2}$Ru$_{2}$O$_{7}$. \cite{Yos1}
With increasing $x$ across the M/NM boundary, $\gamma$ nearly monotonically increases 
without clear divergent behavior. 
This is in contrast with the behavior in Y$_{2-x}$Bi$_{x}$Ru$_{2}$O$_{7}$, 
for which $\gamma$ takes a sharp maximum at the metal-insulator boundary. \cite{Yos1}
In order to explain the apparent discrepancy between our results and the theory 
based on an $s$-electron \cite{Fuj1}, we should take into account of the role of $5d$ electrons, 
which give rise to more complicated band structure compared with that of $s$ electrons. 

In summary, we have revealed that Y$_{2}$Ir$_{2}$O$_{7}$ is a Mott insulator. 
We have also shown that the density of states at the Fermi level 
rapidly changes with $x$ by the filling control of Y$_{2}$Ir$_{2}$O$_{7}$. 
The magnetic ground state appears to occur concomitantly with non-metallic state. 

\begin{figure}
\leavevmode
\epsfxsize=70mm
\epsfbox{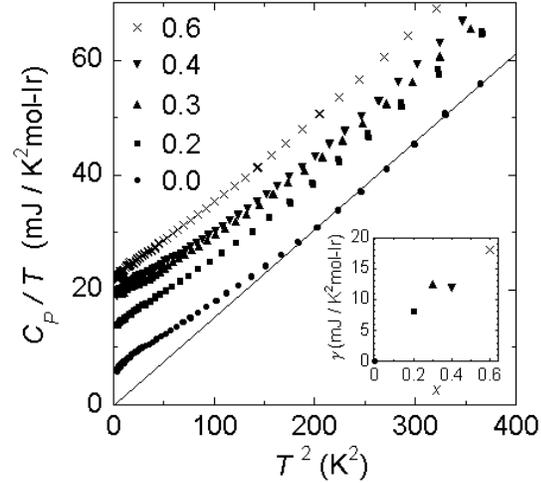}
\caption{Specific heat divided by temperature against squared temperature below 20 K. 
Solid line denotes the quadratic fitting between 14 and 20 K. 
Electronic specific-heat coefficient against the substitution content (inset). }
\label{fig3}
\end{figure}

The authors acknowledge D. Yanagishima for his contributions. 
They would like to thank T. Ishiguro for his support. 
They appreciate stimulating discussion with R. Higashinaka, 
M. Sato, S. Fujimoto and H. Tsunetsugu. 
H. F. acknowledges financial support from JSPS Research Fellowships for 
Young Scientists. 
This work has been supported by the Grant-in-Aid for Scientific Research(S) 
from the Japan Society for Promotion of Science, by the Grant-in-Aid 
for Scientific Research on Priority Area "Novel Quantum Phenomena 
in Transition Metal Oxides" from the Ministry of 
Education, Culture, Sports, Science and Technology, 
and by a Grant from CREST, Japan Science and Technology Corporation.

\end{document}